\documentclass[prd,superscriptaddress,floatfix,nofootinbib,twocolumn]{revtex4-1}

\usepackage{exscale}                  
\usepackage[intlimits]{amsmath}       
\usepackage{amsfonts}
\usepackage{amssymb,amscd}
\usepackage[dvips]{epsfig}                   
\usepackage{graphicx}
\usepackage{array}

\usepackage{subfigure}  

\newcolumntype{L}[1]{>{\raggedright\let\newline\\\arraybackslash\hspace{0pt}}m{#1}}
\newcolumntype{C}[1]{>{\centering\let\newline\\\arraybackslash\hspace{0pt}}m{#1}}
\newcolumntype{R}[1]{>{\raggedleft\let\newline\\\arraybackslash\hspace{0pt}}m{#1}}
\usepackage{color}
\newcommand{\beqa}{\begin{eqnarray}}
\newcommand{\eeqa}{\end{eqnarray}}
\newcommand{\beq}{\begin{equation}}
\newcommand{\eeq}{\end{equation}}
\newcommand{\gS}[1]{#1\!\!\!\!\!\not~}

\newcommand{\qslash}{\gS{q}}
\newcommand{\pslash}{p\!\cdot\!\gamma}

\newcommand{\Tr}{\mathrm{Tr}}

\definecolor{myred}{rgb}{1,0.8,0.8}

\begin{document}
\title{Phase structure of three and four flavor QCD}
\author{Christian~S.~Fischer}\affiliation{Institut f\"ur Theoretische Physik, Justus-Liebig-Universit\"at Gie\ss{}en, Heinrich-Buff-Ring 16, D-35392 Gie\ss{}en, Germany.}
\author{Jan~Luecker}\affiliation{Institut f\"ur Theoretische Physik, Universit\"{a}t Heidelberg, Philosophenweg 16, D-69120 Heidelberg, Germany.}\affiliation{Institut f\"ur Theoretische Physik, Goethe-Universit\"at Frankfurt, Max-von-Laue-stra\ss{}e 1, D-60438 Frankfurt/Main, Germany}
\author{Christian~A.~Welzbacher}\affiliation{Institut f\"ur Theoretische Physik, Justus-Liebig-Universit\"at Gie\ss{}en, Heinrich-Buff-Ring 16, D-35392 Gie\ss{}en, Germany.}
\date{\today}

\begin{abstract}
We investigate the phase structure of QCD at finite temperature and light-quark 
chemical potential. We improve upon earlier results for $N_f=2+1$ dynamical quark 
flavors and investigate the effects of charm quarks in an extension to $N_f=2+1+1$. 
We determine the quark condensate and the Polyakov loop potential using solutions 
of a coupled set of (truncated) Dyson-Schwinger equations for the quark and gluon 
propagators of Landau gauge QCD. At zero chemical potential we find excellent agreement
with results from lattice-QCD. With input fixed from physical observables we find 
only a very small influence of the charm quark on the resulting phase diagram at
finite chemical potential. We discuss the location of the emerging critical
end point and compare with expectations from lattice gauge theory.
\end{abstract} 

\maketitle

\section{Introduction}
Heavy ion collision experiments at BNL, LHC and the future FAIR facility are designed
to probe the quark-gluon plasma (QGP), the state of strongly interacting matter in the 
early Universe a few microseconds after the big bang. In the theoretical description of 
these experiments, in principle two entire quark families have to be taken into account. 
The effect of charm quarks on the equation of state (EoS) is expected to become 
significant at top LHC energies reaching temperatures several times the one of the QCD 
crossover region for light quarks. Hydrodynamical descriptions of the QGP in this 
temperature region therefore need to incorporate the charm quark in their EoS. However, 
even at smaller temperatures at or above the light-quark crossover region, the effects 
of charm quarks on the EoS and the transition temperatures, although predicted to be 
small by perturbation theory \cite{Laine:2006cp}, may not be entirely negligible. 

Precise results from {\it ab initio} calculations on the lattice at zero chemical
potential and physical quark masses are available for $N_f=2+1$ flavors, see e.g. 
\cite{Borsanyi:2010bp,Bazavov:2011nk} and references therein. For the corresponding 
case of $N_f=2+1+1$ flavors only preliminary results for transition temperatures and 
the equation of state using staggered \cite{Bazavov:2013pra,Ratti:2013uta} and Wilson 
type quarks \cite{Burger:2013hia} are available. One of the interesting results of these 
studies is that charm quarks may not be treated in quenched approximation, i.e. the 
backreaction of the charm quarks onto the Yang-Mills sector of the theory is 
quantitatively important \cite{Ratti:2013uta}.

While the lattice results provide excellent guidance for zero baryon chemical potential 
$\mu_B$, the situation becomes much more challenging at $\mu_B \ne 0$ due to the fermion 
sign problem. Various extrapolation methods on the lattice agree with each other for 
$\mu_B/T < 1$, see e.g. \cite{Fodor:2001pe,Bonati:2013tqa,Endrodi:2011gv}. For 
$\mu_B/T > 1$, however, uncertainties accumulate rapidly. Thus other theoretical methods 
are mandatory to complement the lattice calculations.

In this work we use the approach via Dyson-Schwinger equations (DSEs). We update  
previous calculations of the $N_f=2+1$ case and estimate the influence of the charm 
quark on the phase structure of QCD and the location of a putative critical end point (CEP).
One of the advantages of this framework over model treatments is the direct accessibility 
of the Yang-Mills sector of QCD thus rendering a fully dynamical treatment of all members 
of the first two quark families feasible. In order to make the necessary truncations of 
the DSEs well controlled we use constraints such as symmetries and conservation laws as 
well as comparison with corresponding results from lattice calculations when available. 
The goal is then to tighten this control to such an extent that reliable results for large 
chemical potential become feasible. 

The paper is organized as follows. In the next section we explain our truncation scheme,
which is built upon previous work \cite{Fischer:2010fx,Fischer:2012vc,Fischer:2013eca}. 
We use temperature dependent lattice data for the quenched gluon propagator and implement 
the back reaction of the quarks onto the gluons by adding the quark-loop in the gluon-DSE. 
Compared to Ref.~\cite{Fischer:2012vc}, where first results for the $N_f=2+1$ phase diagram 
have been reported, we correct the value of the input up/down quark masses to their physical 
values thus improving the agreement with the lattice results at zero chemical potential. We also 
detail a new procedure to fix the strength of the quark-gluon interaction by solving meson 
Bethe-Salpeter equations in the vacuum according to a novel method introduced in 
Ref.~\cite{Heupel:2014ina}. In Sec.~\ref{sec:results_2p1} we present our updated results
for the $N_f=2+1$ case and compare with the results of lattice QCD where available. In
Sec.~\ref{sec:results_2p1p1} we discuss the effects of the charm quark on the
QCD phase diagram and conclude in Sec.~\ref{sec:sum}.

\section{Order parameters from QCD propagators \label{sec:truncation}}

In order to study the chiral and deconfinement transitions in functional
frameworks such as Dyson-Schwinger equations (DSEs) or the functional
renormalization group (FRG), one needs to specify proper order parameters. For
the chiral transition, the condensate of a quark with flavor $f$,
$\langle\bar{\psi}\psi\rangle_f$, can be extracted from the trace of the quark
propagator $S^f(p)$ via
 \beq \label{eq:condensate}
\langle\bar{\psi}\psi\rangle_f = 
Z_2 Z_m  N_c  T\sum_n\int\frac{d^3p}{(2\pi)^3}\mathrm{Tr}_D\left[S^f(p)\right],
\eeq
where $Z_2$ is the quark wave function renormalization constant, $Z_m$ the
quark mass renormalization constant and $N_c=3$ the number of colors. The sum
is over Matsubara frequencies $\omega_n=\pi T(2n+1)$ and $p =
(\omega_p,\vec{p})$. For all flavors with nonzero bare quark mass the
condensate is quadratically divergent and needs to be regularized. For
dimensional reasons, the divergent part is proportional to the bare quark mass
and therefore the difference 
\begin{equation}
\Delta_{l,s} = \langle\bar\psi\psi\rangle_l - \frac{m_l}{m_s}\langle\bar\psi\psi\rangle_s\,,
\label{eq:cond_renorm}
\end{equation}
fulfils this purpose: the divergent part of the light-quark condensate, $l \in \{u,d\}$, 
is canceled by the divergent part in the strange quark condensate. At physical quark masses 
and small chemical potential, the chiral transition is a crossover, leading to ambiguities
in the definition of a pseudocritical temperature. In this work we use the maximum of 
the chiral susceptibility
\beq \label{eq:chisusz}
\chi_{\langle\bar{\psi}\psi\rangle} = \frac{\partial \langle\bar{\psi}\psi\rangle_l}{\partial m_{l}}\,,
\eeq
as well as the inflection point of the condensate, i.e. the maximum of 
$\frac{\partial \langle\bar{\psi}\psi\rangle_l}{\partial T}$.

The deconfinement transition has been studied with functional methods via the dressed 
Polyakov loop \cite{Fischer:2009wc,Braun:2009gm,Fischer:2010fx,Fischer:2011mz,Fischer:2012vc}, 
the Polyakov loop potential \cite{Braun:2010,Fister:2013bh,Fischer:2013eca} and the analytic 
properties of the quark propagator \cite{MuellerSpec,ThreePolRoberts,Qin:2013ufa}. In this 
work we use the Polyakov loop potential to determine the deconfinement transition at zero 
and finite chemical potential.

\begin{figure}[t]
\includegraphics[width=0.45\textwidth]{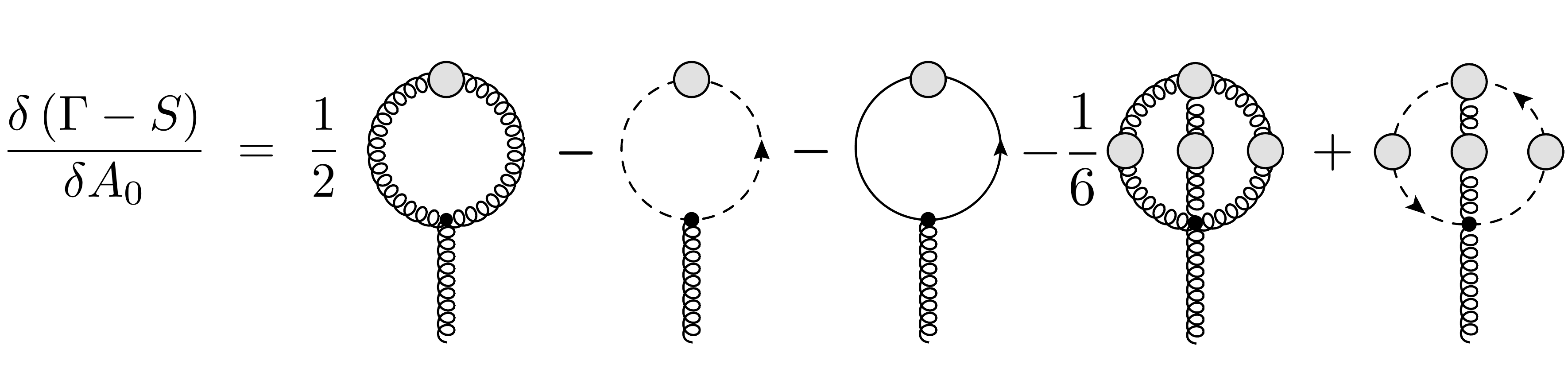}
\caption{The DSE for a background field $\bar{A}_0$. \label{fig:DSE-A}}
\end{figure}

In \cite{Fister:2013bh} the DSE for a background field $\bar{A}_0=\langle
A_0\rangle$ has been introduced; see Fig.~\ref{fig:DSE-A}.  Upon integration
this DSE yields the potential of the background field, which can be connected
to the Polyakov loop by

\beq \label{eq:defPL} L[\langle A_0\rangle] = \frac{1}{N_c} \Tr e^{igA_0/T} \ge
\langle L[ A_0]\rangle.  \eeq
That is, the Polyakov loop evaluated for the background field is an upper bound
for the Polyakov loop expectation value.
If we drop the two-loop diagrams (the last two in Fig.~\ref{fig:DSE-A}), we are
able to obtain this potential solely from the QCD propagators.  This has been
used in \cite{Fischer:2013eca} for the first time for unquenched QCD and at
finite chemical potential. In the same work, it has also been shown that the
deconfinement transition temperature agrees with that obtained from the 
dressed Polyakov loop. Given this agreement, we use the minimum of the Polyakov
loop potential in the approximations discussed in \cite{Fister:2013bh,Fischer:2013eca}
as an order parameter for confinement here.

In order to determine these order parameters we need to specify the propagators of QCD,
i.e. the gluon, ghost and quark propagators. To this end we use a combination of 
lattice methods and solutions from Dyson-Schwinger equations.  

\subsection{Quark and gluon DSEs}

The quark and gluon propagators at finite temperature $T$ and quark-chemical potential 
$\mu$ are given by
\begin{eqnarray}
S^{-1}(p) &=& i\tilde\omega_n\gamma_4C(p)+i\vec{p}\vec{\gamma}A(p)+B(p)\,,\nonumber\\ \label{eq:qProp}\\
D_{\mu\nu}(p) &=& P_{\mu\nu}^L(p)\frac{Z^L(p)}{p^2} + P_{\mu\nu}^T(p)\frac{Z^T(p)}{p^2}\,,
\end{eqnarray}
with momentum $p=(\omega_n,\vec{p})$, $\omega_n=\pi T(2n+1)$ for fermions, 
$\omega_n=\pi T 2n$ for bosons and we use the abbreviation $\tilde\omega_n=\omega_n+i\mu$. 
The projectors $P_{\mu\nu}^{L,T}$ are longitudinal (L) and transversal (T) with respect 
to the heat bath and given by
\begin{eqnarray}
P_{\mu\nu}^T &=& \left(1-\delta_{\mu 4}\right)\left(1-\delta_{\nu 4}\right)\left(\delta_{\mu\nu}-\frac{p_\mu p_\nu}{\vec{p}^{\,2}}\right), \label{eq:projT} \\
P_{\mu\nu}^L &=& P_{\mu\nu} - P_{\mu\nu}^T \,. \label{eq:projL}
\end{eqnarray}

\begin{figure}[t]
\includegraphics[width=0.45\textwidth]{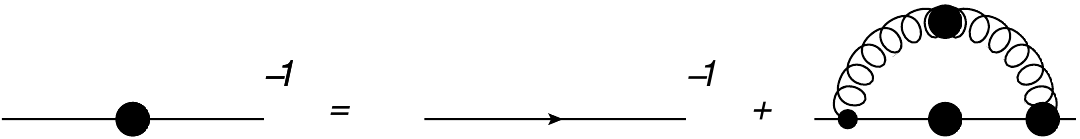}
\caption{The DSE for the quark propagator. Large blobs denote dressed 
propagators and vertices. \label{fig:quarkDSE}}
\includegraphics[width=0.45\textwidth]{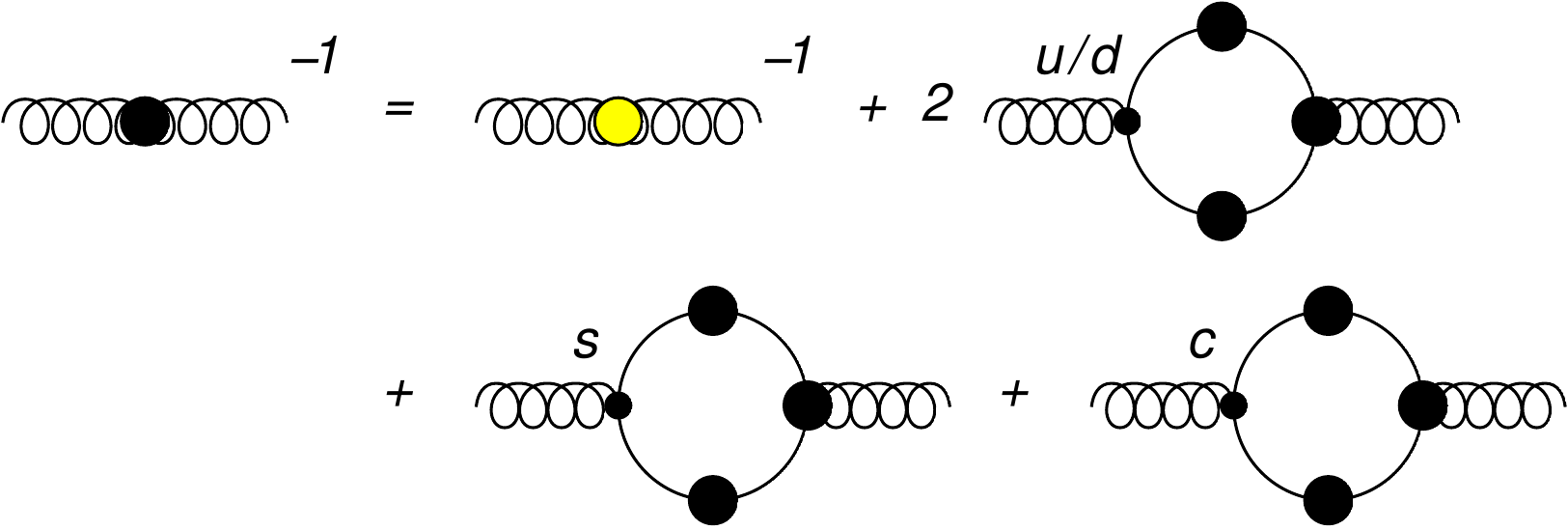}
\caption{The truncated gluon DSE for $N_f=2+1+1$ QCD. The yellow dot denotes the quenched (lattice) propagator. \label{fig:apprGluonDSE}}
\end{figure}

The DSE for the quark propagator is shown diagrammatically in Fig.~\ref{fig:quarkDSE}.
In order to self-consistently solve this equation, we need to specify the fully dressed 
gluon propagator and quark-gluon vertex. Model calculations \cite{Qin:2010nq,Wang:2014yla} 
often use simple Ansatze for the gluon propagator that do not take into account the
proper temperature and flavor dependence of the gluon self-energy. We prefer to include 
these important effects by taking the Yang-Mills sector of QCD into account and calculating 
the backreaction of the quarks onto the gluon explicitly. This framework has been gradually 
evolved from the quenched case, $N_f=0$ \cite{Fischer:2009gk,Fischer:2010fx}, to two flavor
QCD \cite{Fischer:2011mz,Fischer:2011pk,Fischer:2012vc} and recently to $N_f=2+1$ 
\cite{Fischer:2012vc}. It is extended here to include effects of the charm quark.
We believe such an approach has two distinct advantages over simple modeling. On the one
hand it allows us to trace the effects of quark masses and flavors as exposed in the Columbia
plot explicitly in the functional framework. On the other hand, it serves to take into
account the effects of chemical potential on the gluon section of QCD explicitly,
thereby rendering results at finite chemical potential more reliable. 
Furthermore, in our approach we have access to all fundamental degrees of freedom of QCD,
i.e. quark, gluon and ghost propagators, at all values of $T$ and $\mu$.
This allows for the calculation of the Polyakov loop potential, see Fig.~\ref{fig:DSE-A},
and in principle of the full effective action via the functional flow equation.

\renewcommand{\arraystretch}{1.3}
\begin{table*}
\begin{tabular}{|c||C{1.1cm}|C{1.1cm}|C{1.1cm}||C{1.1cm}|C{1.1cm}|C{1.1cm}|C{1.1cm}|C{1.1cm}|}
\hline
Set & $m_{l}$ & $m_{s}$ & $d_1$ &  $m_{\pi}$ & $m_{K}$ & $f_{\pi}$ & $f_{K}$ \\ \hline \hline
$A_{2+1}$& 0.8 & 21.6 & 8.05 & 107 & 405 & 107 & 123   \\ \hline
$B_{2+1}$& 1.32 & 34.1 & 6.8 & 135 & 497 & 94 & 115 \\ \hline

\end{tabular}
\caption{Current quark masses and vertex parameter $d_1$ as well as 
resulting mesonic properties in the vacuum for $N_f = 2+1$.
	The vertex strength $d_1$ is given in GeV$^2$, the other values in MeV.}
\label{tab:para2p1}
\begin{tabular}{|c||C{1.1cm}|C{1.1cm}|C{1.1cm}|C{1.1cm}||C{1.1cm}|C{1.1cm}|C{1.2cm}|C{1.1cm}|C{1.1cm}|C{1.1cm}|}
\hline
Set & $m_{l}$ & $m_{s}$ & $m_{c}$ & $d_1$ &  $m_{\pi}$ & $m_{K}$ & $m_{\eta_c}$ & $f_{\pi}$ & $f_{K}$ & $f_{\eta_c}$ \\ \hline \hline
$A_{2+1+1}$ & 0.8 & 21.6 & 300.0 & 8.05 & 109 & 412 & 2,364 & 95 & 113 & 270 \\ \hline
$B_{2+1+1}$ & 1.23 & 31.6 & 440.0 & 7.6 & 135 & 497 & 2,982 & 94 & 117 & 309 \\ \hline

\end{tabular}
\caption{Current quark masses and vertex parameter $d_1$ as well as resulting mesonic properties in the vacuum for $N_f = 2+1+1$.
	The vertex strength $d_1$ is given in GeV$^2$, the other values in MeV.}
\label{tab:paraNf2p1p1}
\end{table*}

The key idea of our truncation is to replace the Yang-Mills self-energies of the gluon 
DSE with lattice data for the quenched propagator. The resulting gluon DSE is shown in 
Fig.~\ref{fig:apprGluonDSE} for $2+1+1$ flavors. This approximation misses unquenching 
effects in the Yang-Mills self-energies. 
At zero temperature, the effects of this approximation can be explicitly determined
using the framework of Ref.~\cite{Fischer:2003rp}; it is below the five percent level.
We will later on justify this approximation further, by comparing the resulting unquenched
gluon propagator with corresponding lattice results for $N_f=2$.

With the quenched lattice input, the resulting DSEs for the quark and gluon propagators read
\begin{widetext}
\begin{eqnarray}
\left[S^{f}(p)\right]^{-1} &= Z^f_{2}\left[S_0^{f}(p)\right]^{-1} 
+ C_{F}\,Z^f_{1F} \, g^2 \,T\sum_n \int\frac{d^3l}{(2\pi)^3}\, 
\gamma_\mu \,S^f(l)\, \Gamma^f_\nu(l,p;q)\, D_{\mu\nu}(q), \nonumber\\ \label{DSEs} \\
D_{\mu\nu}^{-1}(p) &= \left[D_{\mu\nu}^{qu.}(p)\right]^{-1} - \sum_{f}^{N_f}\,Z_{1F}^f\,\frac{g^2}{2}\,
T\sum_n \int\frac{d^3l}{(2\pi)^3}\, \Tr\left[ \gamma_\mu \,S^{f}(l)\, \Gamma^f_\nu(l,q;p)\,
 S^{f}(q)\right], \nonumber
\end{eqnarray}
\end{widetext}
where $q=(p-l)$, $S^f$ is the quark propagator for one specific flavor $f \in \{u,d,s,c\}$, 
$C_F=\frac{N_C^2-1}{2N_C}$ is the Casimir operator and $\Gamma_\nu$ the dressed quark-gluon vertex.
The vertex and wave function renormalization constants are denoted by $Z_{1F}$ and $Z_2$; for
the running coupling we use $\alpha=g^2/(4\pi)=0.3$.
The remaining quantity to be determined is the dressed quark-gluon vertex $\Gamma_\nu$. 
Here we use the same construction as in previous works (see e.g. \cite{Fischer:2012vc}),
which utilizes the first term of the Ball-Chiu vertex, satisfying the Abelian Ward-Takahashi 
identity (WTI), multiplied with an infrared enhanced function $\Gamma(p^2,k^2,q^2)$ that accounts 
for the non-Abelian dressing effects and the correct ultraviolet running of the vertex.
See Appendix \ref{app:vertex} for more details on our vertex construction.
The resulting expression reads
\begin{widetext}
\begin{eqnarray}
\Gamma_\mu^f(l,p;q) &=& \gamma_\mu\cdot\Gamma(l^2,p^2,q^2) \cdot 
\left(\delta_{\mu,4}\frac{C^f(l)+C^f(p)}{2} + \delta_{\mu,i}\frac{A^f(l)+A^f(p)}{2} \right), \label{vertex1}\\ 
\Gamma(l^2,p^2,q^2) &=& \frac{d_1}{d_2+x} \!
 + \!\frac{x}{\Lambda^2+x}
\left(\frac{\beta_0 \alpha(\mu)\ln[x/\Lambda^2+1]}{4\pi}\right)^{2\delta} \label{vertex2}
\end{eqnarray}
\end{widetext}
where $l$ and $p$ are fermionic momenta and $q$ is the gluon momentum. The dressing functions
$A^f$ and $C^f$ of the quark propagators appearing in Eq.(\ref{vertex1}) introduce a temperature,
chemical potential and quark mass/flavor dependence of the vertex along the WTI. The second term in 
Eq.(\ref{vertex2}) ensures together with the gluon dressing functions the correct logarithmic 
running of the loops in the quark and gluon-DSE. Both scales $\Lambda$ and $d_2$ are fixed 
such that the vertex matches the corresponding scales in the gluon lattice data. In 
\cite{Fischer:2010fx} these have been determined to $d_2 = 0.5$ GeV$^2$ and 
$\Lambda = 1.4$ GeV. The anomalous dimension is $\delta=\frac{-9Nc}{44N_c - 8N_f}$ 
and $\beta_0=\frac{11N_c-2N_f}{3}$. The only free parameter of the 
interaction is the vertex strength $d_1$, which will be discussed below.

We identify the squared momentum variable $x$ with the gluon momentum $q^2$ in the quark DSE 
and with the sum of the two squared quark momenta $l^2+p^2$ in the quark loop. This different 
treatment of the momentum dependence is necessary to maintain multiplicative renormalizability 
of the gluon-DSE \cite{Fischer:2003rp}.
The renormalization procedure of the gluon DSE has
been discussed in detail in \cite{Fischer:2012vc}.\footnote{Here we only mention that transverse
projection in the gluon DSE instead of the Brown-Pennington projection realized in 
\cite{Fischer:2012vc} does not lead to different results but only to a small shift in the
vertex strength parameter $d_1$.}

In general, the different quark flavors are coupled via the DSE of the gluon propagator. 
This leads to a reaction of the strange quark condensate to the chiral transition, as has 
been shown in \cite{Fischer:2012vc}. At the same time, the heavy strange and charm 
quarks influence the light quarks and thus allow for a study of their influence on the
phase diagram in the first place, as already discussed above. In principle, however, 
further coupling effects arise in the DSE for the quark-gluon vertex, which are not 
covered by our truncation scheme. These effects are $1/N_c$-suppressed and may be small.
Nevertheless they should be explored in future work. 

In order to fix the vertex strength $d_1$ as well as the light, strange and
charm quark masses, we follow two strategies: the first one (setup
$A_{N_f}$ in the following) is to reproduce the condensate from lattice QCD as
a function of temperature at $\mu=0$ for $N_f=2+1$ flavors (similar to the
previous work Ref.~\cite{Fischer:2012vc}) and to add a charm
quark without changing $d_1$. The second one (setup
$B_{N_f}$ in the following) is to obtain the masses and decay constants for the
pseudoscalar $\pi$, $K$ and $\eta_c$ mesons in the vacuum. This is done for
$N_f=2+1$ and $N_f=2+1+1$ flavors separately. We will see below, that these two procedures lead
to slightly different results, which may be associated with the systematic error
of our approach.

For the setups $B_{2+1}$ and $B_{2+1+1}$ we need to solve the Bethe-Salpeter equations 
for pseudoscalar mesons in the truncation scheme discussed above. Since our quark-gluon
vertex contains the quark dressing functions $A$ and $C$ depending on the quark momenta,
this cannot be performed with the widely used rainbow-ladder kernel, but requires a more
advanced treatment. In the previous work Ref.~\cite{Fischer:2012vc} the Gell-Mann--Oakes--Renner 
relation has been used to fix the value of the light-quark mass $m_l$ and the ratio 
$m_s/m_l = 27$ subsequently determined the strange quark mass $m_s$. Recent progress in 
the construction of Bethe-Salpeter kernels \cite{Heupel:2014ina} allows us now to solve 
the full Bethe-Salpeter equation including the Ball-Chiu vertex construction. For all 
technical details in this respect we refer the reader to Ref.~\cite{Heupel:2014ina}.

Note that in general the parameter $d_1$ could depend on the quark flavor as
well; see \cite{Williams:2014iea} for an explicit calculation of the vertex
strength for different quark masses. Especially for the charm quark one might
expect a significantly reduced infrared strength of the vertex. We checked the
influence of the reduction of $d_1$ for the charm quark by a factor of 2,
motivated by \cite{Williams:2014iea}. This
leads to marginal changes of our results which are within our estimated
precision. Therefore we keep $d_1$ flavor independent.

\subsection{Quark masses and strength of the quark-gluon interaction \label{sec:BSEfixing}}

\begin{figure}[t]
\includegraphics[width=0.4\textwidth]{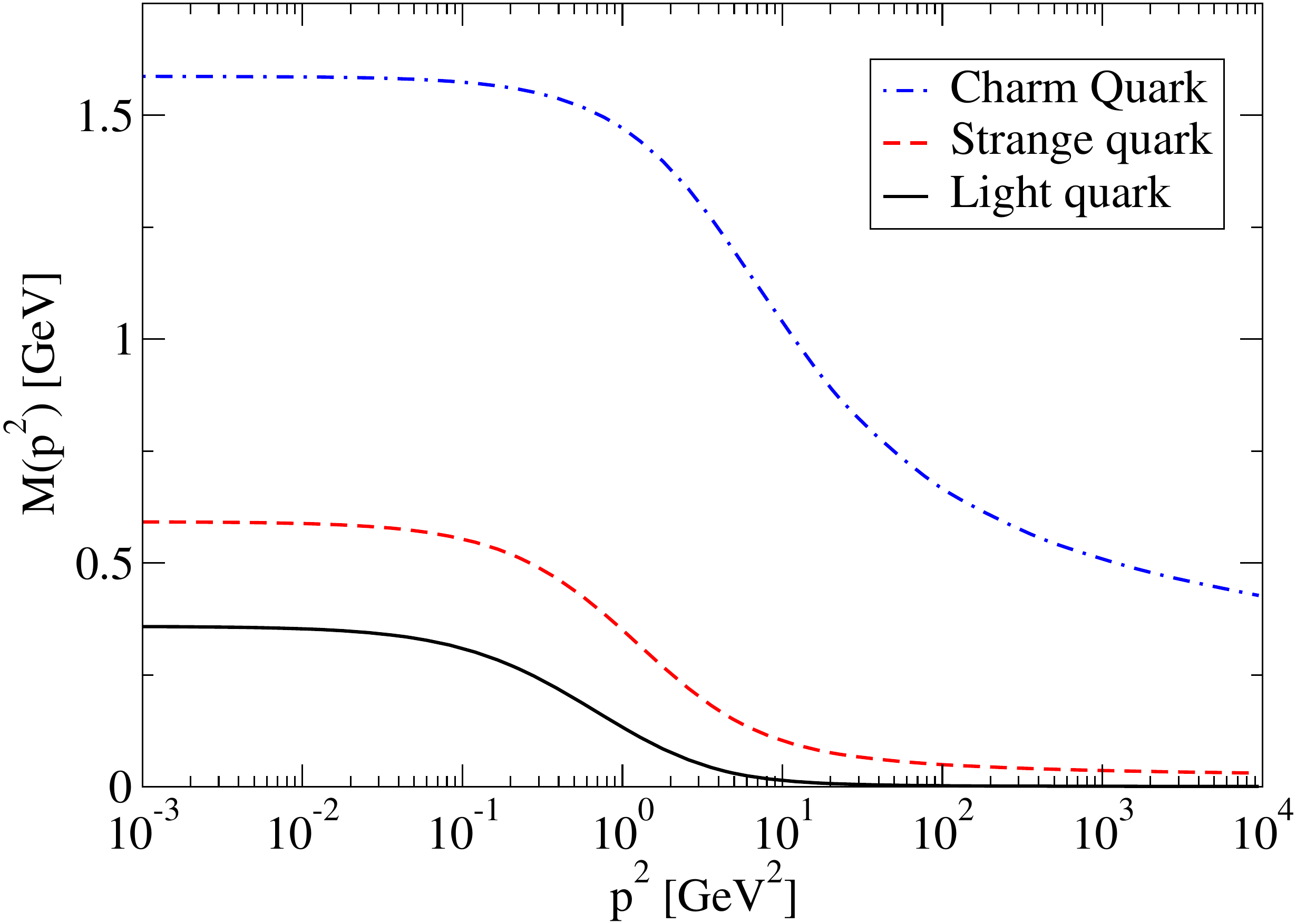}
\caption{The vacuum quark mass function $M(p^2)=B(p^2)/A(p^2)$ for light, strange and charm quarks in parameter set $B_{2+1+1}$. \label{fig:massFunction}}
\end{figure}

In Tables \ref{tab:para2p1} and \ref{tab:paraNf2p1p1} we summarize the resulting sets 
of quark masses and vertex strengths $d_1$ for this work. The sets $B_{2+1}$ and $B_{2+1+1}$ 
are obtained by solving the Bethe-Salpeter equation as described above, while in set 
$A_{2+1}$ we choose $d_1$ to match the chiral transition temperature at zero chemical
potential for $N_f=2+1$ taken from the lattice \cite{Borsanyi:2010bp}, similar
as in Ref.~\cite{Fischer:2012vc}. For set $A_{2+1+1}$ we merely added an additional 
charm quark. As can be 
seen in Tables \ref{tab:para2p1} and \ref{tab:paraNf2p1p1}, the ratio of 
the strange quark to light-quark mass resulting from the fixing procedure using the 
Bethe-Salpeter equation is $m_s/m_l \approx 26$, which is very close to the value 
$m_s/m_l \approx 27$ used in \cite{Fischer:2012vc}. The ratio of the charm to strange 
quark mass for set $B_{2+1+1}$ is $m_c / m_s \approx 14$. These results are within the 
same ballpark as results from lattice calculations, see Refs.~\cite{Carrasco:2013laa,Bazavov:2013nfa}.
Note, however, that the quark masses are renormalization point and scheme dependent,
a direct quantitative comparison is therefore not easily possible. For sets A we choose the ratio 
$m_s/m_l = 27$ and inherited $m_c / m_s \approx 14$ from set $B_{2+1+1}$ for comparability.

The quark masses depend on the renormalization point $\zeta$, $m=m(\zeta)$. We choose a 
rather large $\zeta=80$ GeV here, in order to be sufficiently far in the perturbative regime.
The seemingly small charm quark masses of $m_c=300$ MeV and $m_c=440$ MeV are a result of 
this large renormalization 
point. In Fig.~\ref{fig:massFunction} we show the quark mass function $M(p^2)=B(p^2)/A(p^2)$
for all considered quark flavors to illustrate its momentum dependence.

\section{Results \label{sec:results}}

\subsection{Unquenched gluon propagator}

\begin{figure}[h!]
\centering
\subfigure[Longitudinal part \label{fig:sfig1}]{\includegraphics[width=.85\linewidth]{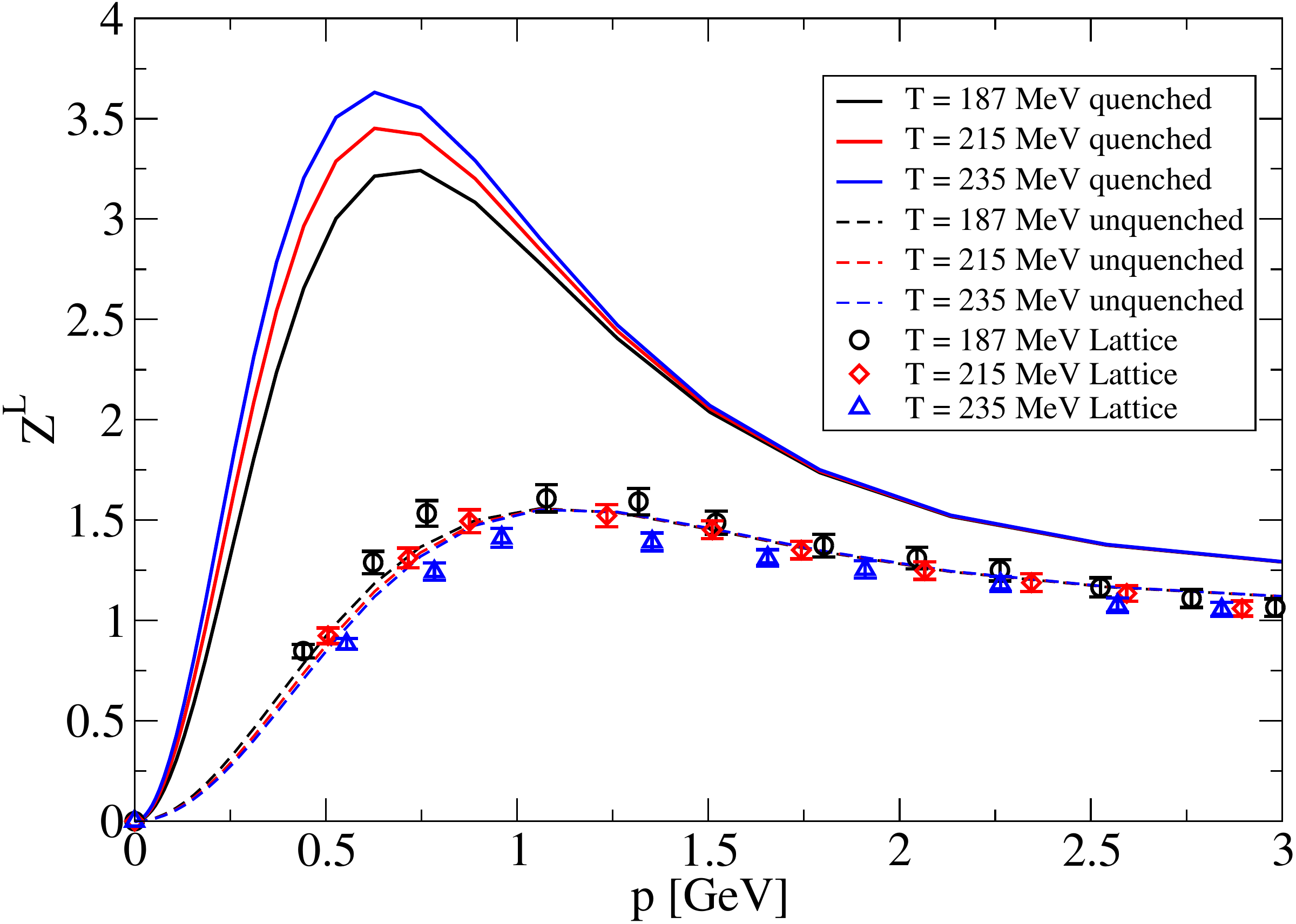}}\\
\subfigure[Transversal part \label{fig:sfig2}]{\includegraphics[width=.85\linewidth]{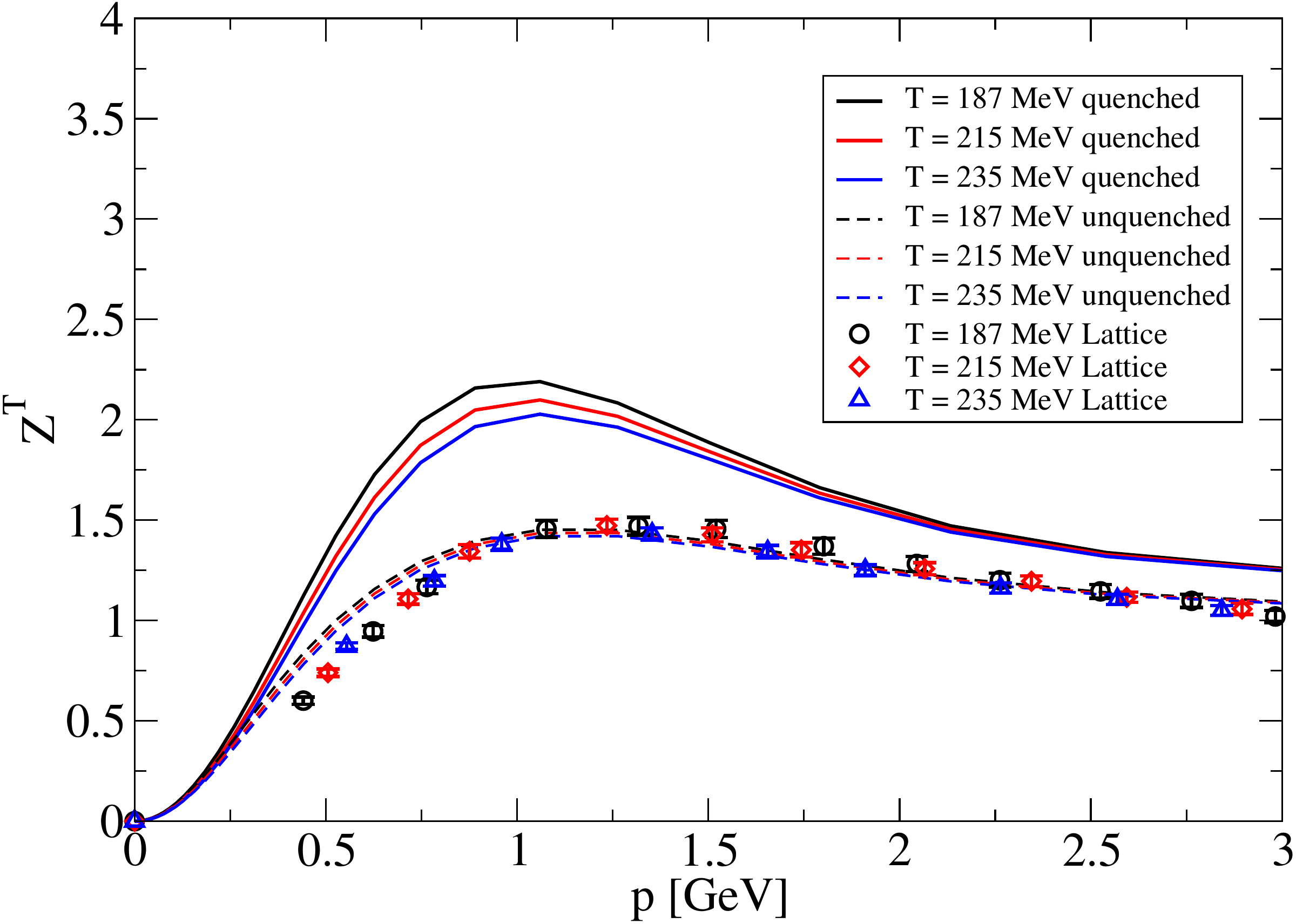}}
\caption{\label{fig:GluonCompLat} Comparison of gluon dressing functions for $N_f$=2 in 
our DSE approach \cite{Fischer:2013eca} with lattice data \cite{Aouane:2012bk}. All results
have been evaluated at a pion mass of $m_\pi = 316$ MeV.}
\end{figure}

First results for the unquenched gluon propagator for $N_f=2$ and $N_f=2+1$ at finite 
temperature have been reported in Ref.~\cite{Fischer:2012vc} and compared with the 
lattice results of Ref.~\cite{Aouane:2012bk}. 
Here we give an update of this comparison with an adjusted vertex strength $d_1$ and light-quark mass $m_l$ in the manner of sets B.
For $N_f=2$ we fixed the parameter $d_1$ and the light-quark mass using its
Bethe-Salpeter equation (BSE) to reproduce
the vacuum mass and decay constant of the pion. After that we increased the light quark mass until the result of the BSE matched a pion 
mass of $m_\pi = 316$ MeV, leading to $d_1 = 5.3$ GeV$^2$ and $m_l=$7.95 MeV.

\begin{figure*}[t]
\centering
\subfigure[Longitudinal part \label{fig:sfig5}]{\includegraphics[width=.45\linewidth]{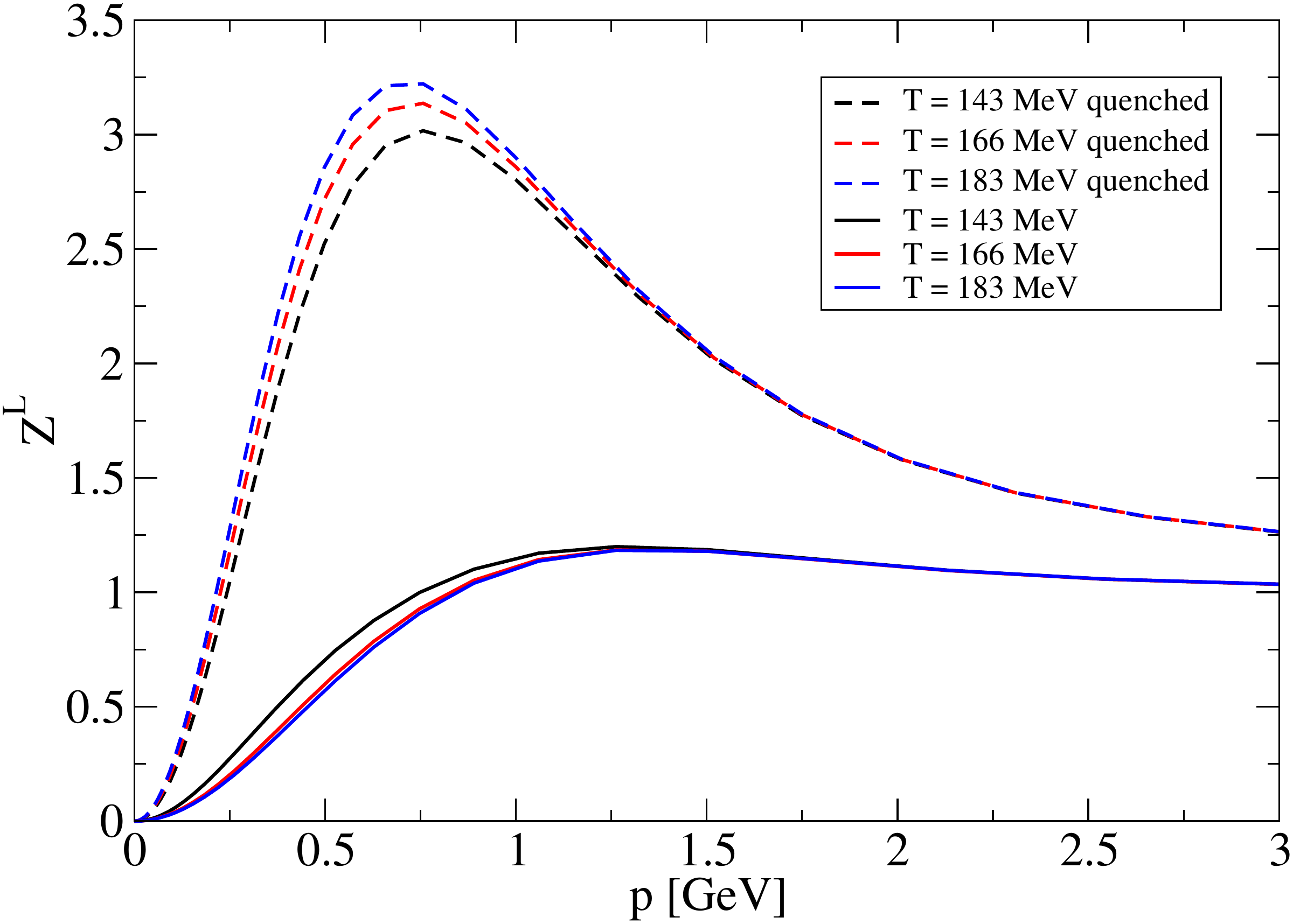}}
\hfill
\subfigure[Transversal part \label{fig:sfig6}]{\includegraphics[width=.45\linewidth]{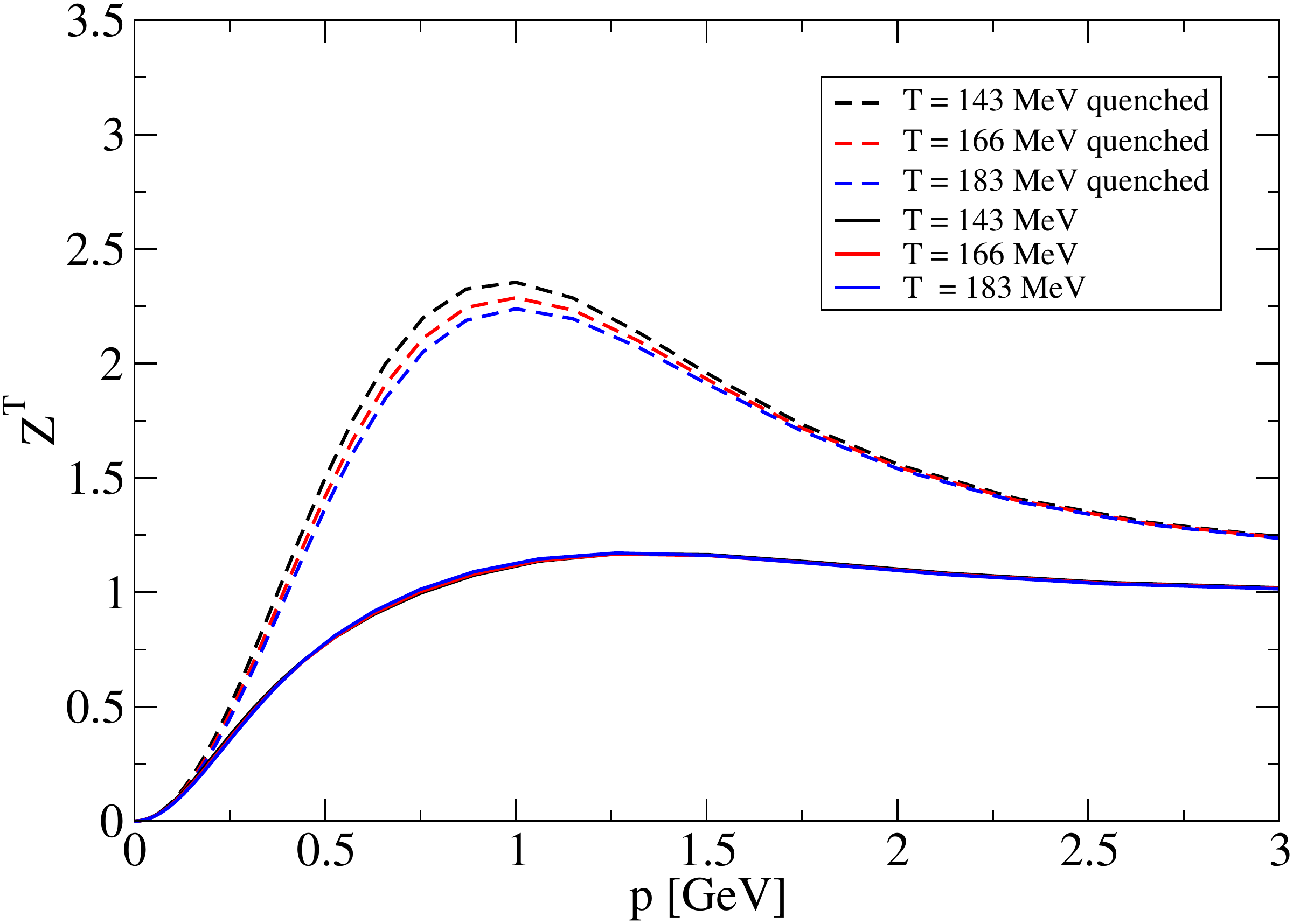}}
\caption{\label{fig:GluonNf3} Gluon dressing functions for $N_f=2+1$ and physical quark masses
(set $A_{2+1}$) at three different temperatures}
\subfigure[Longitudinal part \label{fig:sfig3}]{\includegraphics[width=.45\linewidth]{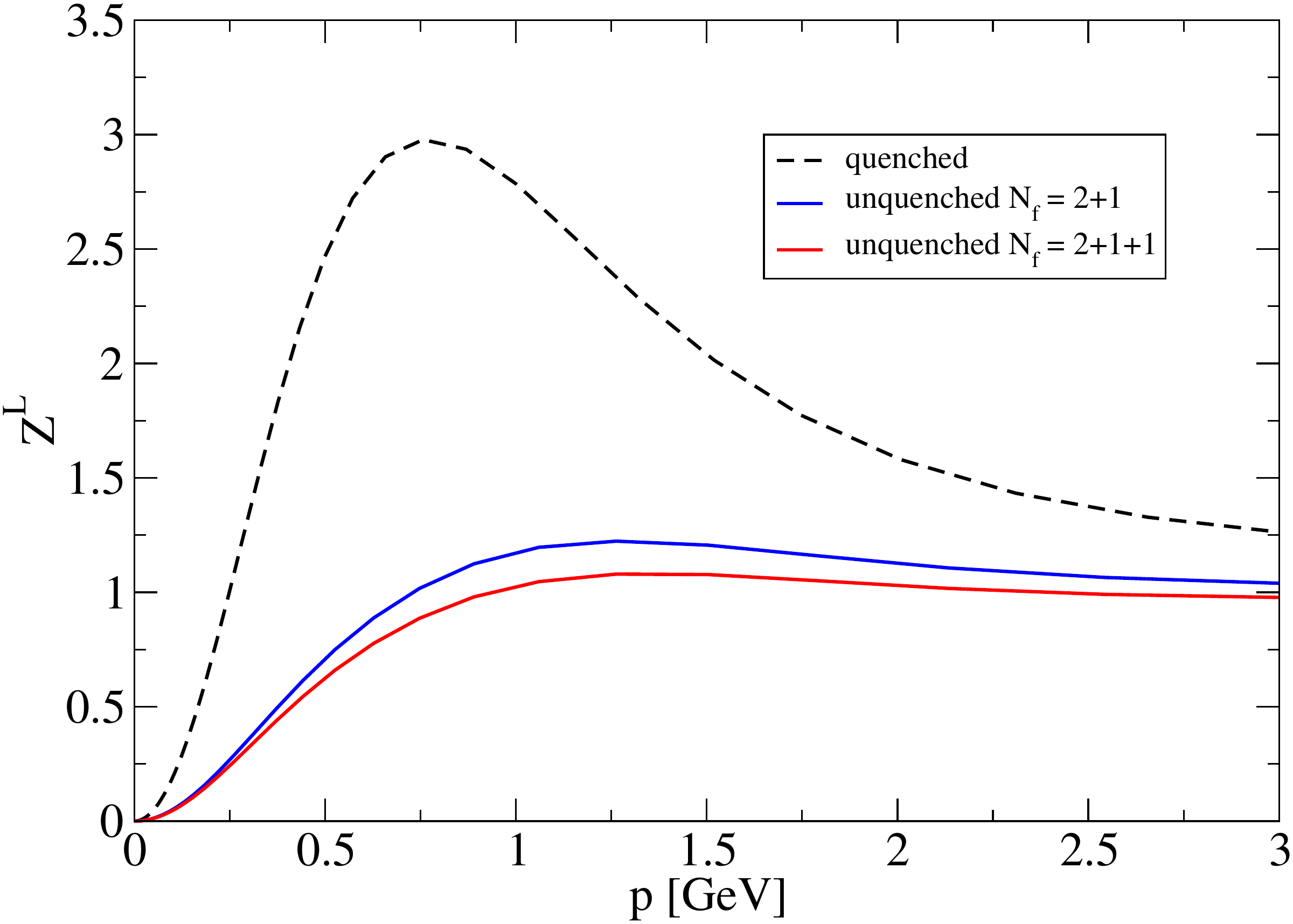}}
\hfill
\subfigure[Screening mass \label{fig:sfig4}]{\includegraphics[width=.45\linewidth]{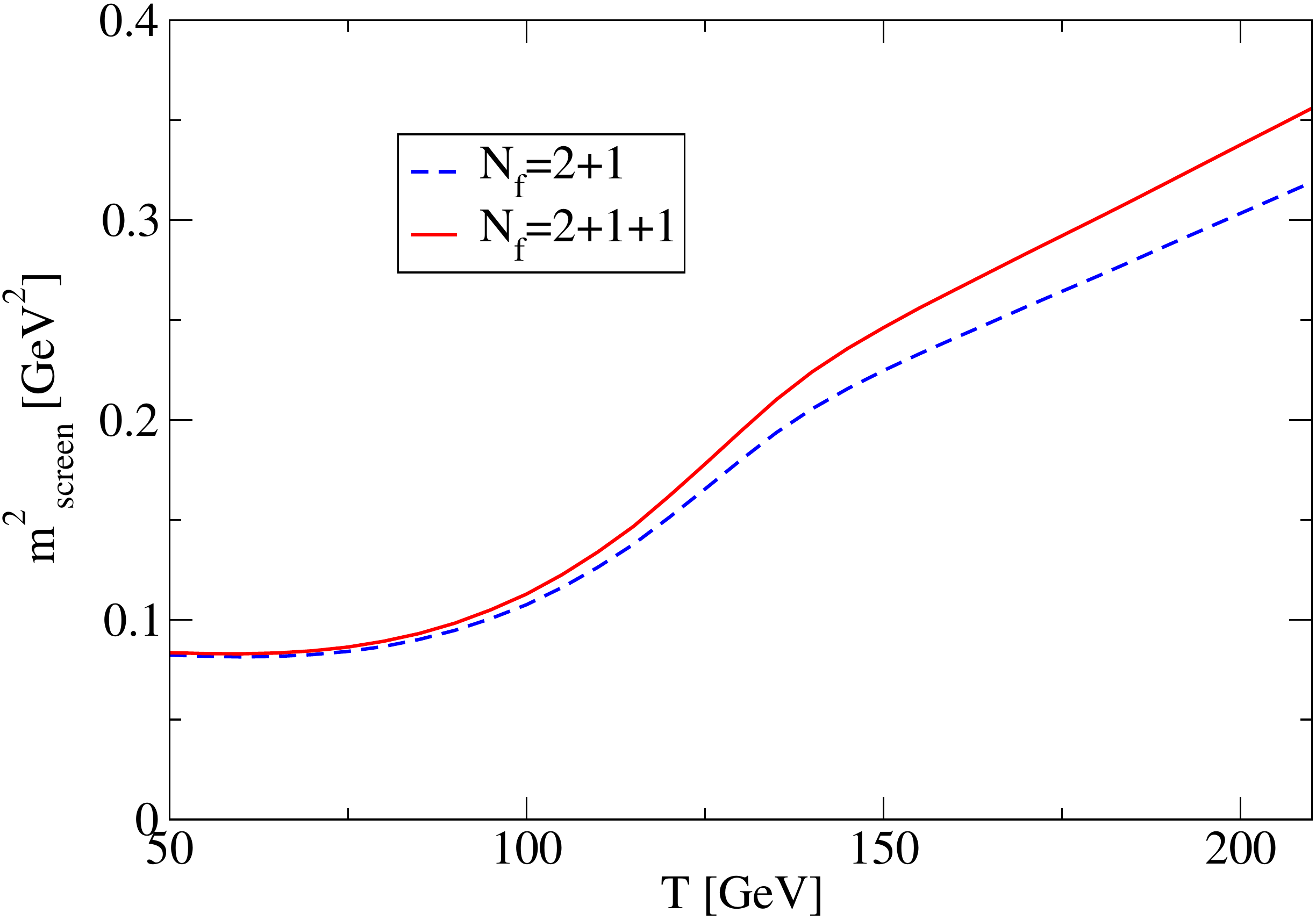}}
\caption{\label{fig:GluonScreenMass} Left: Longitudinal gluon dressing function for $N_f=2+1$ and $N_f=2+1+1$ 
physical quark masses (sets $B_{2+1}$, $B_{2+1+1}$) at $T \approx 135$ MeV. Right: Corresponding gluon 
screening mass as a function of temperature.}
\end{figure*}

The results for the gluon dressing functions are shown in Fig.~\ref{fig:GluonCompLat}. We find large
unquenching effects in both, the magnetic and electric part of the gluon propagator. These
affect the momentum dependence of the gluon with a large reduction of the size of the
bump in the nonperturbative moment region. Furthermore, the quark loop effects 
even invert the temperature dependence of the electric gluon dressing function $Z_L$: for the temperatures 
shown the bump in the quenched dressing function increases with $T$ \cite{Fischer:2010fx}, 
whereas it decreases in the unquenched case. This prediction of the DSE-framework has 
been verified by the lattice calculations \cite{Aouane:2012bk}. In general, the 
quantitative agreement between the two approaches is very good and justifies to some 
extent our truncation scheme.

In Fig.~\ref{fig:GluonNf3} we display our results for the gluon dressing 
functions with physical up/down and strange quark masses (set $A_{2+1}$).
Compared to  Fig.~\ref{fig:GluonCompLat} we find a further reduction of the bump in the 
dressing function due to the increased screening effects of the lighter quarks.
The results of Fig.~\ref{fig:GluonNf3} are our prediction for the unquenched
gluon at physical quark masses and should be checked by future lattice calculations.

In order to gauge the effects of the charm quark on the gluon we compare the
$N_f=2+1$ and $N_f=2+1+1$ result (sets $B_{2+1}$ and $B_{2+1+1}$) in 
Fig.~\ref{fig:sfig3} for $T = 135$ MeV, close to the 
pseudocritical temperature of this parameter set. We observe a further reduction of the
dressing functions up to about 15 percent close to the bump and the expected change
in the large momentum behavior due to different anomalous dimensions. Similar changes
can be observed in the transverse gluon dressing function, not shown in the figure.
A good measure for the effects in the deep infrared is the change of the screening 
mass 
\begin{equation}
m^{2}_{screen} = \left[\frac{p^2}{Z^L(p^2)}\right]_{p^2 \rightarrow 0}
\end{equation}
 in the electric gluon,
shown in Fig.~\ref{fig:sfig4}.
We observe that for small temperatures, where the quark contribution to the
screening mass is small, charm quarks have a negligible effect. At larger temperatures
the effect of the charm quarks is of the order of ten percent, growing to a factor of
$4/3$ for asymptotic temperatures.
In Sec.~\ref{sec:results_2p1p1} we will discuss the consequences 
of these changes for the chiral and deconfinement transition. 

\begin{figure*}[t]
\includegraphics[width=0.45\textwidth]{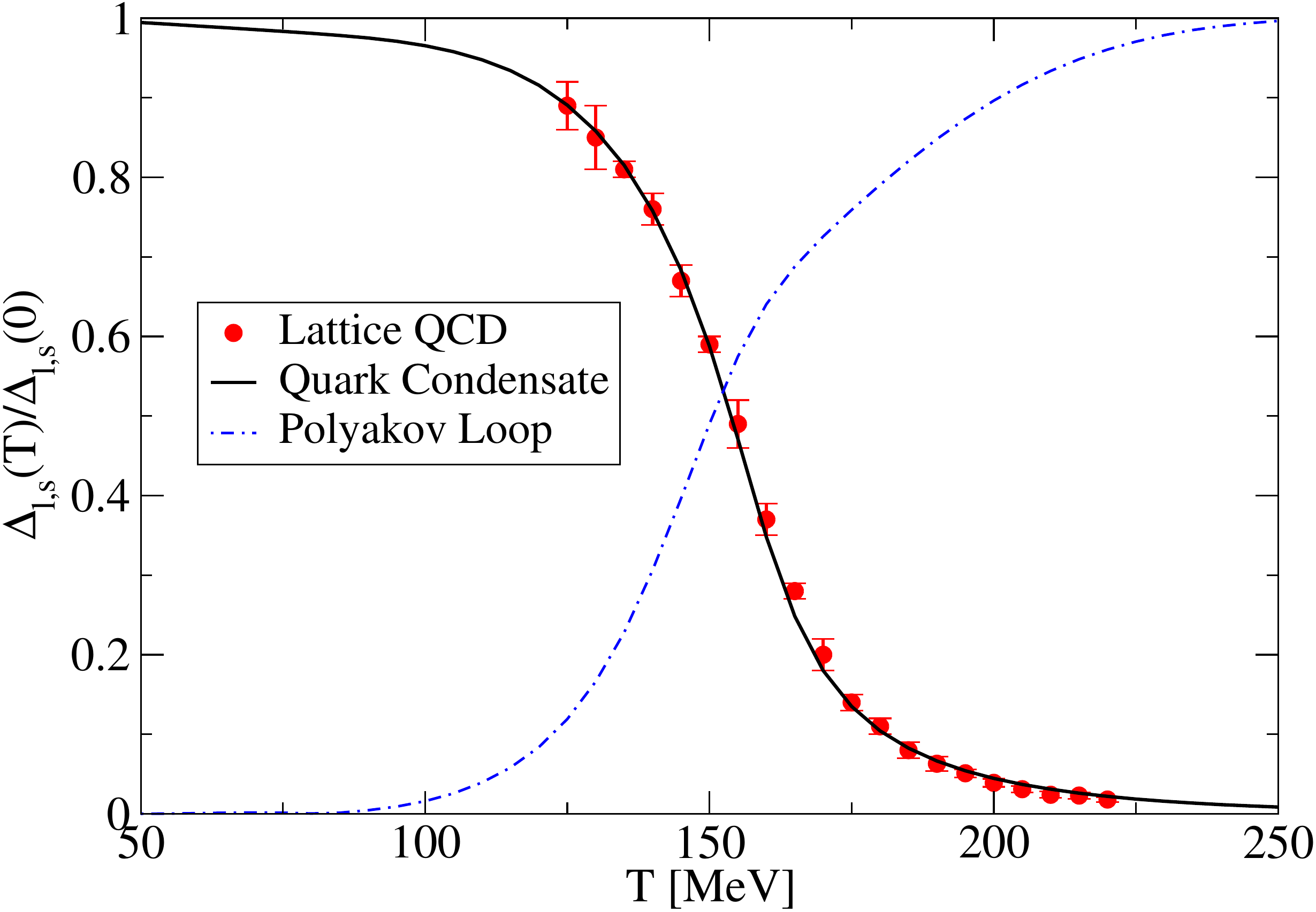}\hfill
\includegraphics[width=0.45\textwidth]{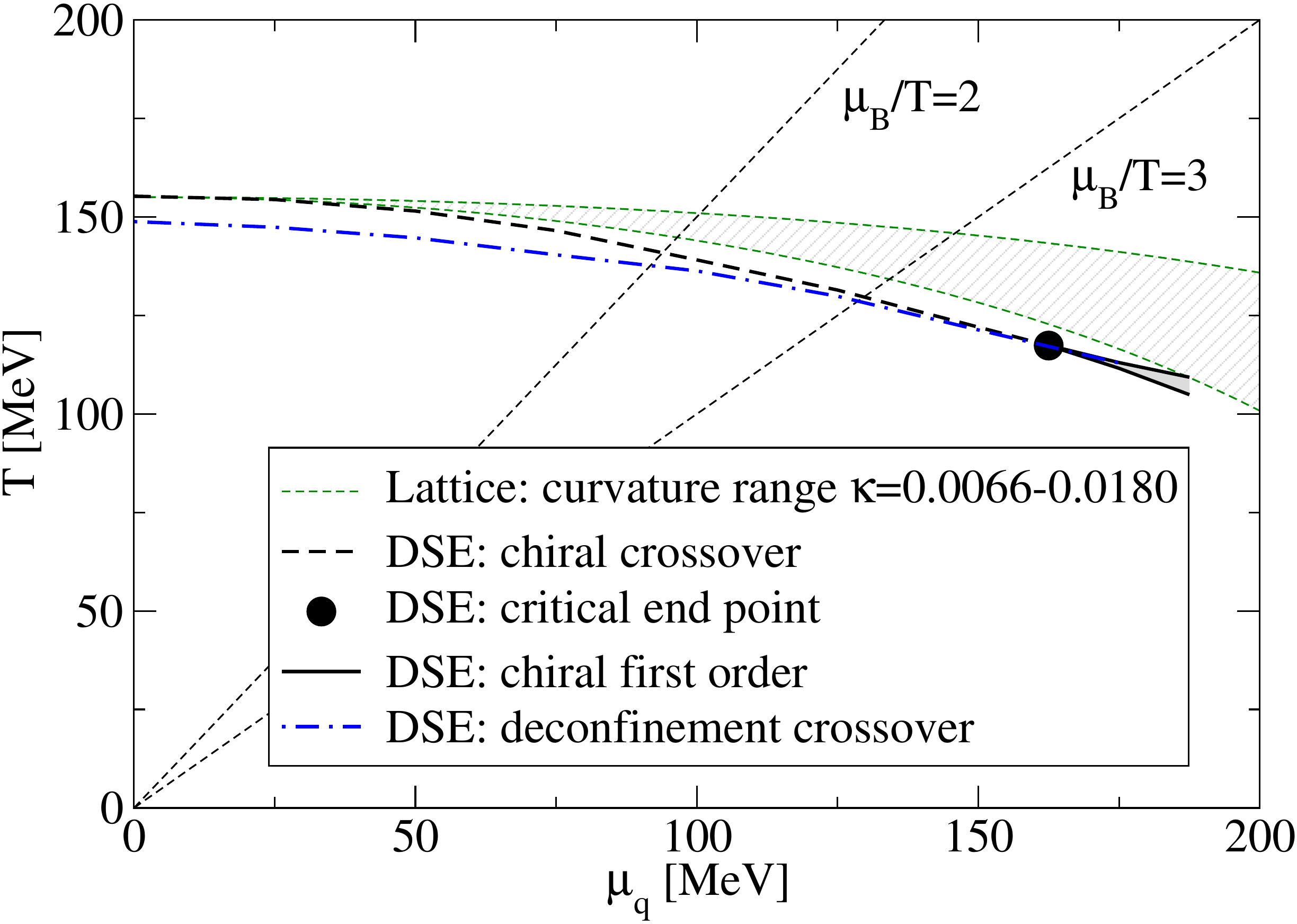}
\caption{Left diagram: regularized chiral condensate and the Polyakov loop 
for $N_f=2+1$ quark flavors (set $A_{2+1}$) as a function of temperature at zero quark chemical 
potential $\mu_q = 0$. Right diagram: phase diagram for $N_f=2+1$ quark flavors.
Shown are our results for set $A_{2+1}$ together with an extrapolation of a range of curvatures for 
the chiral transition extracted at imaginary and small chemical potential from 
different lattice groups \cite{Endrodi:2011gv,Kaczmarek:2011zz,Cea:2014xva}. 
\label{fig:regcond}}
\end{figure*}

\subsection{The QCD transition for $N_f=2+1$ quark flavors \label{sec:results_2p1}}

Before we discuss the influence of the charm quark, we present the updated results
for the chiral and deconfinement transition with $N_f=2+1$ physical up/down 
and strange quark masses. In the left diagram of Fig.~\ref{fig:regcond} we display 
the regularized quark condensate as well as the Polyakov loop as a function 
of temperature at zero chemical potential. Compared to the results reported in
Ref.~\cite{Fischer:2012vc} we have corrected a factor of two in the determination of the
up/down quark mass. As a result, we find much better agreement with the lattice data
especially in the temperature region above the chiral restoration, where the effects
of the explicit chiral symmetry breaking are most notable. As explained above, the
strength of the quark-gluon interaction, controlled by the parameter $d_1$, has been
adjusted in our calculation such that the transition temperature of the lattice data 
is reproduced. The nontrivial result of our calculation is the perfect agreement of 
the steepness of the chiral transition with the lattice result 
(see Ref.\cite{Herbst:2013ufa} for a corresponding result in the Polyakov loop quark-meson model). This agreement 
together with the agreement for the unquenched gluon discussed above, shows that
our truncation scheme works very well at zero chemical potential. The resulting transition
temperatures from the chiral susceptibility and the the inflection point of the light-quark
condensate for set $A_{2+1}$ are
\begin{eqnarray}
\left.T_c\right|_{\frac{d\langle\bar\psi\psi\rangle}{dm}} &= 160.2 \,\mbox{MeV}\,, \nonumber \\ 
\left.T_c\right|_{\frac{d\langle\bar\psi\psi\rangle}{dT}} &= 155.6 \,\mbox{MeV}.
\end{eqnarray}

Our results for the QCD phase diagram at finite chemical quark potential are shown
in the right diagram of Fig.~\ref{fig:regcond}. We extracted the (pseudo-) critical 
temperature of the chiral transition from the inflection point, 
Eq.~(\ref{eq:chisusz}). Compared to Ref.~\cite{Fischer:2012vc} we only find small
corrections due to the corrected light-quark masses. The chiral
crossover, displayed by the dashed black line, becomes ever steeper with increasing
chemical potential and turns into a CEP at 
\begin{equation}
(T^c,\mu_q^c)=(115,168) \,\mbox{MeV}. 
\end{equation}

The deconfinement transition line is determined via the minimum of the Polyakov loop potential 
\cite{Fischer:2013eca}. The relatively large difference of chiral and deconfinement
transition temperatures at small $\mu$ is in part an effect of using the inflection point 
for the Polyakov loop potential on the one hand and on the other hand applying the 
maximum of the susceptibility in the chiral transition. In \cite{Fischer:2013eca} 
the inflection point has been used for both order parameters, yielding closer critical 
temperatures. At large chemical potential, the deconfinement transition line meets 
the chiral one at the CEP. To better guide the eye,
we have also marked lines with ratios of baryon chemical potential over temperature
$\mu_B/T = 2$ and $\mu_B/T = 3$, further underlining that the CEP occurs at rather large
chemical potential.

\begin{figure*}[t]
\includegraphics[width=0.60\textwidth]{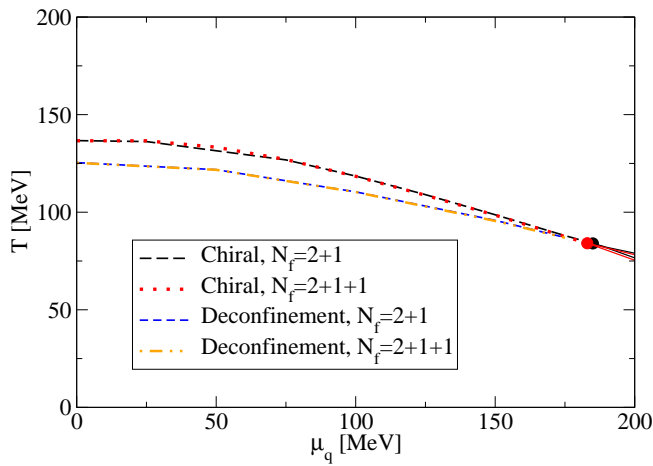}
\caption{Phase diagram for QCD with $N_f=2+1$ and $N_f=2+1+1$ flavors (sets $B_{2+1}$ and $B_{2+1+1}$).
In both cases, the interaction strength and quark masses are fixed to 
reproduce vacuum physics. \label{fig:phasediagram}}
\end{figure*}

In order to gauge the quality of our result, a couple of comments are in order. First, note
that in our representation of the quark-gluon interaction via Eqs.~(\ref{vertex1}),(\ref{vertex2})
no explicit effects of the backcoupling of mesons and baryons onto the quark propagator
have been included. In principle, such effects are encoded in the details of the
Dyson-Schwinger equation for the vertex and have been made explicit in Ref.~\cite{Fischer:2007ze}. 
In the vacuum, these effects are included implicitly within the form of our 
vertex Ansatz Eqs.~(\ref{vertex1}),(\ref{vertex2}) and the choice of the parameter $d_1$.
At $\mu = 0$ and finite quark mass these effects appear to be unimportant,
as demonstrated by the agreement with the quark condensate from the lattice, as discussed 
above. At finite chemical potential, however, meson and baryon effects in the
vertex introduce an additional temperature and chemical potential dependence of
the quark-gluon interaction on top of the ones already covered by our Ansatz.
For example, in PQM studies mesonic fluctuations have been found to have a
large effect on the position of the CEP; see \cite{Herbst:2010rf,Herbst:2013ail}. 
Furthermore baryon effects, which are certainly
important in the low temperature, large chemical potential region in the
vicinity of the nuclear liquid-gas transition, may be crucial. Whether these
contributions have a large impact on the location (or even on the very
existence) of the CEP needs to be checked in future work.

It is tempting to compare our result for the chiral transition line with the lattice 
extrapolations. To this end we also display in Fig.~\ref{fig:regcond} the extrapolation 
of the curvature of the chiral transition line from $N_f=2+1$ 
lattice results of different 
groups at imaginary and zero chemical potential \cite{Endrodi:2011gv,Kaczmarek:2011zz,Cea:2014xva}
into the real chemical potential region (for recent results with $N_f=2$ see
\cite{Cea:2012ev}). Overall, the agreement between the lattice extrapolation 
and our DSE results is quite satisfactory.
However, we wish to add that a similar caveat as for the DSEs may also apply to the lattice
extrapolation. Since the effects of baryons on the chiral transition are small at small
chemical potential they are not reflected in the curvature extracted from the lattice results
and therefore it remains an open question, to what extent an extrapolation to large chemical
potential can be trusted. Thus the close agreement of both approaches, although interesting,
may very well not be the final word.

\subsection{Including the charm quark: $N_f=2+1+1$ \label{sec:results_2p1p1}}

As explained above, there are two possibilities of how the charm quark 
can be added to our truncation. First the interaction strength $d_1$ from
the $N_f=2+1$ case can be kept fixed and the charm quark merely be added
to the system. Such a procedure leads to a reduction of the chiral
and deconfinement transition temperatures of $\Delta T \approx 23$ MeV 
for all values of the chemical potential. This procedure, however, does
not reflect the physics of the charm quark properly and leads to a gross
overestimation of its effects. Instead, we follow another procedure and
determine the vertex strength $d_1$ as well as the current quark masses
using input from hadron physics at $T=0$, as described in Sec.~\ref{sec:BSEfixing} 
(setups $B_{2+1}$ and $B_{2+1+1}$). Our result for 
the corresponding phase diagram is shown in Fig.~\ref{fig:phasediagram},
where we used the chiral susceptibility to determine the chiral 
transition. Note that the transition temperature for $N_f=2+1$ is lower 
by about $\Delta T \approx 20$ MeV compared to the one discussed in 
Sec.~\ref{sec:results_2p1}, due to the different procedure of fixing the interaction 
strength. One can 
view this difference as the systematic uncertainty of our truncation scheme.
The influence of the charm quark on the chiral and deconfinement transition 
is almost negligible apart from a small shift of the critical end point
towards smaller chemical potential. This confirms our expectations from the
unquenched gluon propagator. Despite a 15 percent effect at large momenta,
the low momentum change in the propagator is small enough not to affect the
chiral and deconfinement properties of the theory. Ultimately, this is tied to 
the fact that thermal effects in the charm quarks are small due to its large
mass. Since the vacuum effects of the charm have been absorbed in the readjustment
of the vertex strength from setup $B_{2+1}$ to $B_{2+1+1}$, the overall effect of
the charm quark is almost negligible. We expect to see a similar behavior
in corresponding lattice calculations. 
In future studies it may therefore be sufficient to include only the dynamics of 
light and strange quarks.

\section{Summary and conclusions \label{sec:sum}}

We solved the coupled system of Dyson-Schwinger equations for the quark and
gluon propagators for $N_f=2+1$ and $N_f=2+1+1$ quark flavors using a truncation 
scheme that takes quark fluctuations in the gluon propagator into account. For
the Yang-Mills part of the gluon self-energy we employed temperature dependent
lattice data as input. For the quark-gluon interaction we used a form that 
incorporates temperature and chemical potential effects according to (the 
leading part of) a Ward identity. Furthermore we adapted the infrared strength of 
this interaction such that the chiral transition temperature of lattice gauge 
theory is reproduced. As a highly nontrivial result of our approximation scheme 
we obtained excellent agreement for both, the detailed shape of the chiral transition
as well as the momentum and temperature dependence of the resulting unquenched
gluon propagator at zero chemical potential. 

From the quark and gluon propagators we determined the chiral susceptibility as well
as the Polyakov loop potential as order parameters for the chiral and deconfinement
transition. In the resulting QCD phase diagram we identified a chiral critical end 
point at large chemical potential $(T^c,\mu_q^c)=(115,168)$ MeV, where $\mu_B/T > 3$.
Whether our approximation scheme is still trustable at this point remains to be
investigated in future work, where we plan to take meson and baryon effects in the
quark-gluon interaction explicitly into account.

We also evaluated, for the first time, the effects of a fourth flavor on the
chiral critical end point. This affects the light-quark condensate indirectly, via 
the back coupling of the charm quark onto the unquenched gluon propagator. We 
established that this effect is sizeable for the momentum dependence of the gluon
at intermediate and large momenta. However, for low momenta and for the temperatures
relevant for the chiral transition, the gluon propagator remains essentially unchanged
such that the chiral transition temperature remains the same within our numerical 
uncertainty of 1--2 MeV. Finite chemical potential does not change this situation such
that the location of the critical end point is hardly affected by the charm.
Therefore we established, for the first time in a nonperturbative approach, that 
charm quarks do not affect the QCD phase diagram.

\vspace*{5mm}
{\bf ACKNOWLEDGEMENTS}\\
We thank Jan Pawlowski, Bernd-Jochen Schaefer and Lorenz von Smekal for fruitful 
discussions and Walter Heupel for providing the code for the BSE calculations.
This work has been supported by the Helmholtz International Center for 
FAIR within the LOEWE program of the State of Hesse.

\appendix

\section{The quark-gluon vertex \label{app:vertex}}

Here we explain in more detail our construction for the quark-gluon
vertex we use in our approach. In general, this vertex satisfies a
Slavnov-Taylor identity (STI) \cite{Marciano:1977su} given by 
\begin{eqnarray}
i\: k_\mu \: \Gamma_\mu(q,k) =&& G(k^2) \times \\
&& \times [S^{-1}(p) H(p,q) - \bar{H}(q,p) \: S^{-1}(q)],\nonumber
\label{quark-gluon-STI}
\end{eqnarray}
where $G(k^2)$ denotes the dressing function of the ghost propagator and $H(q,p)$ 
a ghost-quark scattering kernel with "conjugate" $\bar{H}$. The momenta of the two 
quarks are given by $p,q$ and $k=p-q$ is the corresponding gluon momentum. Since the
nonperturbative behavior of $H(p,q)$ and its conjugate is currently unknown,
there is no exact solution of this identity available (see, however, 
\cite{Aguilar:2013ac,Rojas:2013tza} for recent progress in this direction). 
A valid strategy to work along this identity at least approximately is to start 
with the corresponding Abelian identity, where $G=H=\bar{H}=1$. Using the requirement 
of regularity at zero gluon momentum this identity has been solved by the
Ball-Chiu vertex 
\cite{Ball:1980ay}. At zero temperature and chemical potential it is given by
\begin{eqnarray}
\Gamma_\nu^{BC}(p,q,k) &=& \frac{A(p^2)+A(q^2)}{2} \gamma_\nu \\
&& + i \frac{B(p^2)-B(q^2)}{p^2-q^2} (p+q)_\nu
\nonumber\\
&& + \frac{A(p^2)-A(q^2)}{2(p^2-q^2)} (\pslash+\qslash)(p+q)_\nu \nonumber
\label{vertex_BC}
\end{eqnarray}
For the present calculation we retain the leading $\gamma_\mu$-part of this
construction, generalized to finite temperature. This is the content of 
Eq.~(\ref{vertex1}) in the main text. 

Comparing the structure of the WTI with the STI one is able to infer additional
information on the vertex \cite{Fischer:2003rp}. First, there is the factor 
$G(k^2)$ on the right-hand side of the STI. The ghost dressing function at finite
temperature is known from lattice calculations \cite{Fischer:2010fx} and exhibits
an (almost) temperature independent enhancement at infrared momenta. Approximate 
treatments of the ghost-quark scattering kernel at zero temperature show a similar 
enhancement in the infrared \cite{Aguilar:2013ac,Rojas:2013tza}. In the absence of 
more information we approximate the combined effects of the ghost dressing
function and the scattering kernel by a function $\Gamma(k^2)$, Eq.~(\ref{vertex2}),
which is temperature (and chemical potential) independent and a function
of the gluon momentum only. The final construction of Eqs.~(\ref{vertex1}),(\ref{vertex2})
then consists of a factorized non-Abelian part $\Gamma$ and the leading tensor structure 
of the Abelian Ball-Chiu construction.

The infrared effects of the remaining parts of the Ball-Chiu vertex as well as the 
eight transverse parts of the vertex not constrained by the WTI can be thought of as 
absorbed in the 
parameter $d_1$, representing the infrared strength of all components
of the vertex. The resulting dressing function $\Gamma$ represents the generic 
momentum running of the leading dressing functions of the vertex as extracted
from explicit results for the vertex DSE at zero temperature (see \cite{Williams:2014iea} 
and references therein for recent results): these functions run logarithmically at 
large momenta, become comparably large at typical infrared QCD scales and then 
stay constant in the deep infrared. This is the content of Eq.~(\ref{vertex2}). 

Clearly, from a systematic point of view our approximation of the vertex is still 
crude. It contains, however, some important elements which provide some justification
for its use. First, it is correct in the perturbative momentum domain, where the leading 
Ball-Chiu part dominates and the dressing function $\Gamma$ contains the correct running 
of one-loop resumed perturbation theory. Second, it maintains charge conjugation symmetry
required of the full vertex. Third, it contains at least some of the presumed temperature 
and chemical potential dependence of the full vertex via the leading Ball-Chiu term.
Finally, and most important, it provides for results that reproduce and successfully 
predict lattice results for the chiral condensate and the unquenched gluon propagator.
This is detailed in the main body of this work.

\bibliography{./PaperBib}		

\end{document}